\documentclass[ aip,
reprint,%
]{revtex4-2}
\usepackage{graphicx}
\usepackage[svgnames]{xcolor}
\usepackage{amsmath, amssymb}

\newcommand{\mylab}[3]{\raisebox{#2}[0mm][0mm]%
           {\makebox[0mm][l]{\hspace*{#1}{#3}}}}

\newcommand{\opensquare}{\mbox{$\square$}}
\newcommand{\opentriangle}{\mbox{$\vartriangle$}}
\newcommand{\opentriangledown}{\mbox{$\triangledown$}}
\newcommand{\opendiamond}{\mbox{$\lozenge$}}

\newcommand{\opencirc}{\mbox{\Large$\circ$}}
\newcommand{\opensqr}{\mbox{$\square$}}

\newcommand{\dotted}{\protect\mbox{${\mathinner{\cdotp\cdotp\cdotp\cdotp\cdotp\cdotp}}$}}
\newcommand{\dashed}{\protect\mbox{-\ -\ -\ -}}

\def\spacce#1{\hskip #1pt}
\def\drawline#1#2{\raise 2.5pt\vbox{\hrule width #1pt height #2pt}}
\def\solid{\drawline{24}{.5}\nobreak}

\def\bdash{\hbox{\drawline{5.8}{.5}\spacce{2}}}

\def\dashed{\bdash\bdash\bdash\nobreak}

\def\bdot{\hbox{\drawline{1}{.5}\spacce{2}}}

\def\dotted{\hbox{\leaders\bdot\hskip 24pt}\nobreak}

\def\chndot{\hbox%
{\drawline{4.6}{.5}\spacce{2}\drawline{1}{.5}\spacce{2}\drawline{4.6}{.5}\spacce{2}\drawline{1}{.5}\spacce{2}\drawline{4.6}{.5}}\nobreak}

\def\dd{{\, \rm{d}}}
\def\dr{{\rm{d}}}

\def\bra{\langle}
\def\ket{\rangle}
\def\p{\partial}

\def\beq{\begin{equation}}
\def\eeq{\end{equation}}
\def\la{\label}

\def\r#1{(\ref{#1})}

\def\bm{\boldsymbol{m}}

\def\utau{u_\tau}
\def\retau{Re_\tau}

\def\notyet#1{{\color{red}#1}}
\def\notyet#1{{#1}}

\newcommand{\figpath}{./}

\begin{document}
\title{The eddies are attached, but it is all right.}

\author{Javier Jim\'enez}
\email[]{jjsendin@gmail.com}
\affiliation{School of Aeronautics, Universidad Polit\'ecnica de Madrid, 28040 Madrid, Spain}

\date{\today}

\begin{abstract}

The behavior of velocity fluctuations near a wall has long fascinated the turbulence community,
because the prevalent theoretical framework of an attached-eddy hierarchy appears to predict
infinite intensities as the Reynolds number tends to infinity. Although an unbounded infinite limit is not a problem in
itself, it raises the possibility of unfamiliar phenomena when the Reynolds number is large, and has
motivated attempts to avoid it. We review the subject and point to possible pitfalls stemming from
uncritical extrapolation from low Reynolds numbers, or from an over-simplification of the multiscale
nature of turbulence. It is shown that large attached eddies dominate the high-Reynolds-number
regime of the near-wall layer, and that they behave differently from smaller-scale ones. In that
limit, the near-wall layer is controlled by the outer flow, the large-scale fluctuations reduce to
a local modulation of the near-wall flow by a variable friction velocity, and the kinetic-energy peak is
substituted by a deeper structure with a secondary outer maximum. The friction velocity is then not
necessarily the best velocity scale. While the near-wall energy peak probably becomes unbounded in
wall units, it almost surely tends to zero when expressed in terms of the outer driving velocity.
\end{abstract}
  
\maketitle

\section{Introduction}\label{sec:intro}

While there is reasonable agreement that turbulence should become independent of
viscosity in the limit of very large Reynolds numbers \cite{kol41,onsag}, this is not true for
wall-bounded turbulent flows, where even bulk quantities such as the friction factor slowly decay
when the Reynolds number increases \cite{tenn}. The reason is that there is a layer near the wall
where viscosity is always needed to enforce the no-slip boundary boundary condition and, even if
the relative thickness of this layer steadily decreases with the Reynolds number, the classical
argument that leads to the logarithmic velocity profile \cite{millikan38} also implies that the velocity
gradients grow without limit near the wall, and that their contribution to the production and dissipation of the 
turbulence fluctuations cannot automatically be neglected.

As a consequence of this singular behavior, much of the discussion about the high-Reynolds number
limit of wall-bounded turbulence has centered on the intensity of the near-wall peak of the streamwise
velocity fluctuations, $u'_p$, empirically located at a distance $y_p^+\approx 15$ from the wall,
where the `+' superindex denotes `wall' normalization with the friction velocity $u_\tau$ and with the
kinematic viscosity $\nu$. In numerical simulations, its magnitude approximately increases
logarithmically with the friction Reynolds number, $Re_\tau =u_\tau h/\nu$, where $h$ is the flow
thickness \cite{jim18}. The evidence from experiments is more mixed, but not incompatible with
numerics when both are available \cite{deGraaf00,mar:etal:PRF17,jim18}, and a similar logarithmic behavior is
found in the spanwise velocity and in the pressure, although not in the fluctuations of the
wall-normal velocity \cite{jim18}. The implication that the velocity fluctuations become
infinitely strong in the limit of infinitely high Reynolds number has caused some unease, leading to
repeated efforts to avoid it \cite{chen:sree:21,monk:22,hwang24} and to restore the asymptotic
$Re_\tau$-independence of wall turbulence.

This paper briefly reviews those efforts. One of our first conclusions will be that the available
range of experimental and numerical Reynolds numbers is too narrow for an unguided extrapolation to
$Re\to \infty$, and is likely to remain so for some time. The decision between different models
should rather come from theoretical arguments, if possible, and most of the paper deals with how
such theoretical models are constrained by the data. We denote the streamwise, wall-normal and
spanwise directions by $x, y$ and $z$, respectively, and the corresponding velocity components by
$u, v$ and $w$. Capital letters denote $y$-dependent ensemble averages, $\bra\,\ket$, as in the mean
velocity profile, $U(y)$, and lower-case ones are fluctuations with respect to these averages.
Primes refer to root-mean-square (rms) fluctuation intensities.

In principle, an infinitely strong near-wall intensity peak does not present insurmountable
theoretical difficulties. If we assume an approximately constant tangential Reynolds stress
throughout the logarithmic layer, $\bra u v\ket \approx -u_\tau^2$, and a logarithmic profile for
the mean velocity \cite{tenn},
\beq
U^+=A +\kappa^{-1} \log y^+,
\la{eq:loglaw}
\eeq
where $A\approx 5$, and $\kappa\approx 0.4$ is the K\'arm\'an constant,  
the total energy production in the flow is \cite{tenn}
\beq
 -\int_0^h \bra u v\ket\, \p_y U \dd y \approx \int_{\delta_{in}}^{Re_\tau} \frac{u_\tau^3 \dd y^+}{\kappa y^+} 
 \sim \frac{u_\tau^3}{\kappa} \log Re_\tau, 
\la{eq:logprod}
\eeq
which grows without bound with $Re_\tau$. The reason for the singularity is not the lower bound of
the integral in \r{eq:logprod}, which can be regularized by a suitable viscous cut-off, but the
divergence of the integral of the mean shear, $\p_y U\sim O(1/y)$, in its upper limit,
$O(h^+=Re_\tau)$. Unless the logarithmic profile is assumed to cover a vanishingly small fraction of
the flow thickness as $Re_\tau$ increases, Eq. \r{eq:logprod} implies an infinitely large production
of energy, at least when expressed in wall units, that could conceivably leak towards the wall and
result in infinitely strong fluctuations in its neighborhood. Note, in particular, that the bulk
velocity from Eq. \r{eq:loglaw}, which arguably drives the flow, also becomes infinite as
$\retau\to\infty$.

However, it is probably true that such a situation would result in extremely strong fluctuations
near the wall when $\retau$ is large but finite. There is some evidence of strong fluctuations in
the atmospheric boundary layer \cite{met:klew:01,metetal01} but, even in that case, their intensity,
$u'^+_p\approx 4$ at $\retau\sim O(10^6)$, is weak with respect to the mean velocity at the same
distance from the wall $U^+(y_p)\approx 10.5$. In fact, even using the most unfavorable of the
extrapolations discussed below for $u'^+_p$, the intensity of the near-wall fluctuation peak only
becomes comparable to the mean profile for $\retau\approx 10^{74}$, which is well beyond any
reasonable extrapolation from experimental or observational Reynolds numbers. More damaging from the
theoretical point of view is the assumption that the viscosity required to enforce the boundary
conditions remains relevant for infinitely strong velocity structures, and that the associated flow
features remain stable.

A recent survey and discussion of currently popular models can be found in Ref.
\onlinecite{hwang24}, and will not be repeated here. They can broadly be classified in two groups.
The first one are structural models that propose mechanisms for how the flow is organized. The
best-known is the attached-eddy model, first proposed by \textcite{tow:61}, and structurally
developed by many others \cite{per:cho:82,phc86,SmitMcKMar11,DeshMonMar:21}. It is the main support
for the unbounded logarithmic growth of the near-wall intensity, and will be discussed later in more
detail. The models in the second group typically assume a finite intensity asymptote at
$\retau\to\infty$, and search for the most mathematically and physically consistent form of the
defect with respect to this asymptote. The two best-known proposals are a
logarithm \cite{monk:nag:15,monk:22,hwang24},
\beq
u'^2_p(\retau) =u'^2_p(\infty) -O(1/\log \retau),
\la{eq:logdef}
\eeq
 and a power law \cite{chen:sree:21,piroz:24},  
\beq
u'^2_p(\retau) =u'^2_p(\infty) -O(\retau^{-1/4}),
\la{eq:powdef}
\eeq
which are, unfortunately, difficult to distinguish within the available range of Reynolds number. In
this paper, we will mostly restrict ourselves to exploring the consequences of the attached-eddy
model, both from the point of view of the near-wall fluctuations and of its interplay with the
viscous boundary condition.

We will do this by comparing theoretical arguments with a homogeneous set of data presented in
\S\ref{sec:data}. The attached-eddy model and its supporting evidence are discussed in
\S\ref{sec:AE}, and its weak points are discussed in \S\ref{sec:results}, together with possible
solutions. This first part of the paper deals mostly with the fluctuations of the streamwise
velocity component, but \S\ref{sec:w} extends the argument to the spanwise velocity. Conclusions are
summarized in \S\ref{sec:conc}.

\section{The data set}\la{sec:data}
\begin{table}
\caption{\label{tab:data}%
Numerical data sets most often used in the paper. The box dimensions are $L_x\times 2h \times L_z$, and 
the symbols in the rightmost column are used consistently in the figures.}
\begin{ruledtabular}
\begin{tabular}{lcccc}
Reference& $L_x/h$ & $L_z/h$ & $Re_\tau$ & Symbol\\
\hline
\textcite{jc06_vor}      & $8\pi$ &  $3 \pi$ & 180--950 & \opencirc \\ 
\textcite{hoyas06} & $8\pi$ &  $3\pi$ & 2000 &  \opencirc \\ 
\textcite{lozano14}      & $60\pi$ &  $6 \pi$ & 550 & \opensquare \\ 
\textcite{lozano14}      & $2\pi$ &  $\pi$ & 4200 & \opensquare \\ 
Bernardini {\em et al.} \cite{BerPirOrl:14} & $6\pi$  & $2\pi$ & 180--4000 & \opentriangle \\
\textcite{lee:moser:15} & $8\pi$ &  $3 \pi$  & 550--5200 & \opendiamond \\ 
\textcite{hoyas22} & $2\pi$ &  $\pi$ & 10000 & \opentriangledown \\ 
\end{tabular}
\end{ruledtabular}
\end{table}

We will mostly restrict ourselves to numerical doubly periodic pressure-driven turbulent channel
flow between parallel plates separated by $2h$, for which there is a reasonably homogeneous set of
simulations spanning two orders of magnitude in Reynolds number. They are summarized in Tab.
\ref{tab:data}. At the same time, these simulations include enough variety of computational
parameters, mainly in the numerical method and in the size of the computational box, to provide some
safeguard against overfitting our results to one particular technique. Although restricting us in
this way limits our conclusions somewhat, it avoids the scatter due to variable resolution in
experiments, while keeping a range of $\retau$ comparable to what can reliably be measured
experimentally. Moreover, since our discussion will lead us to consider the effect on the near-wall
region of the largest flow scales, the restriction to channels avoids the known differences between
their large scales and those in pipes \cite{ng_11} or in boundary layers
\cite{jim_etal_10,sillero14,monk:nag:15}. Most data sets in table \ref{tab:data} include mean
fluctuation profiles, one- and two-dimensional spectra, and energy balances, and are freely
accessible from the web pages of the different groups.

Spectra and cospectra are used in premultiplied form, $\Phi_{ab} (k_x, y)= k_x \bra \hat a \hat
b\ket$, where $\hat a (k_x)$ is the Fourier coefficient of the expansion of $a(x)$ in terms of the
streamwise wavenumber $k_x$, with similar definitions for spanwise spectra in terms of $k_z$, and for
two-dimensional ones in terms of both wavenumbers. They will usually be expressed as functions of the
wavelengths, $\lambda_j = 2\pi/k_j$.

\begin{figure}
\centerline{%
\hspace{.03\columnwidth}\raisebox{0mm}{\includegraphics[width=.78\columnwidth,clip]%
{\figpath 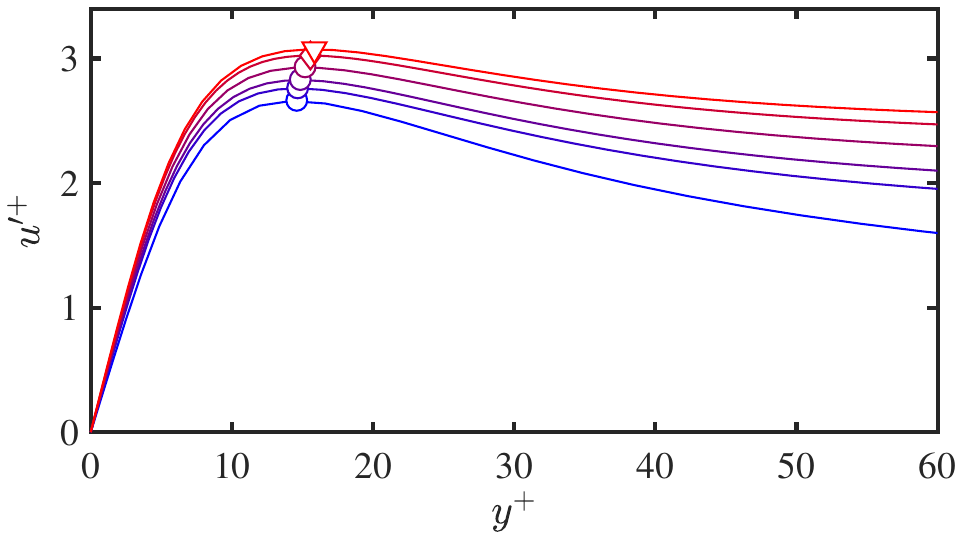}}
\mylab{-.85\columnwidth}{.38\columnwidth}{(a)}%
}%
\centerline{%
\hspace{.03\columnwidth}\raisebox{0mm}{\includegraphics[width=.78\columnwidth,clip]%
{\figpath 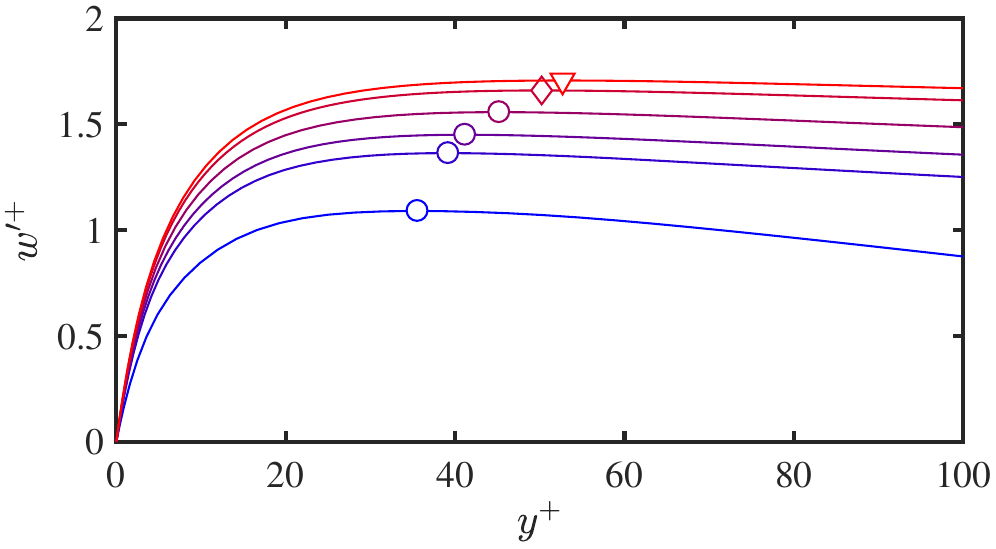}}
\mylab{-.85\columnwidth}{.38\columnwidth}{(b)}%
}%
\centerline{
\raisebox{0mm}{\includegraphics[width=.78\columnwidth,clip]%
{\figpath 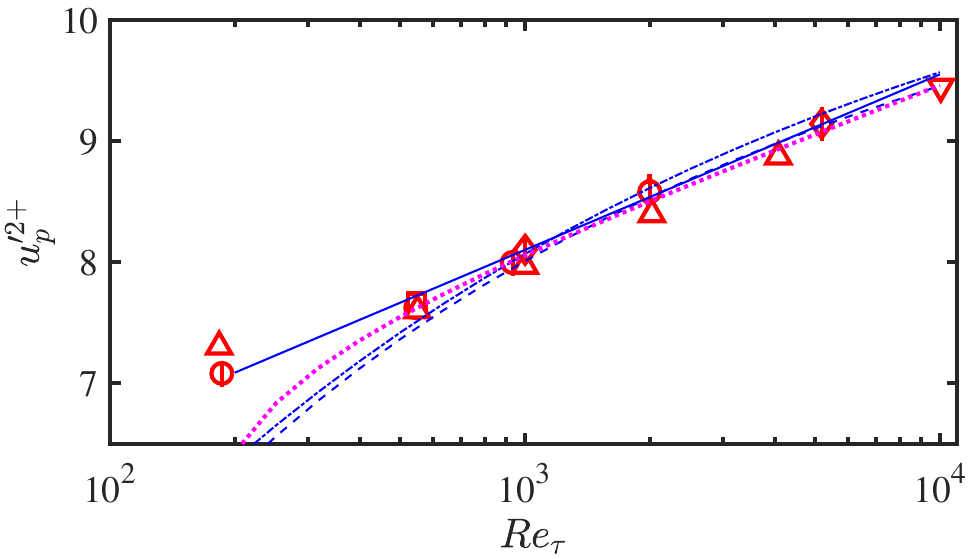}}
\mylab{-.85\columnwidth}{.38\columnwidth}{(c)}%
}%
\centerline{
\raisebox{0mm}{\includegraphics[width=.78\columnwidth,clip]%
{\figpath 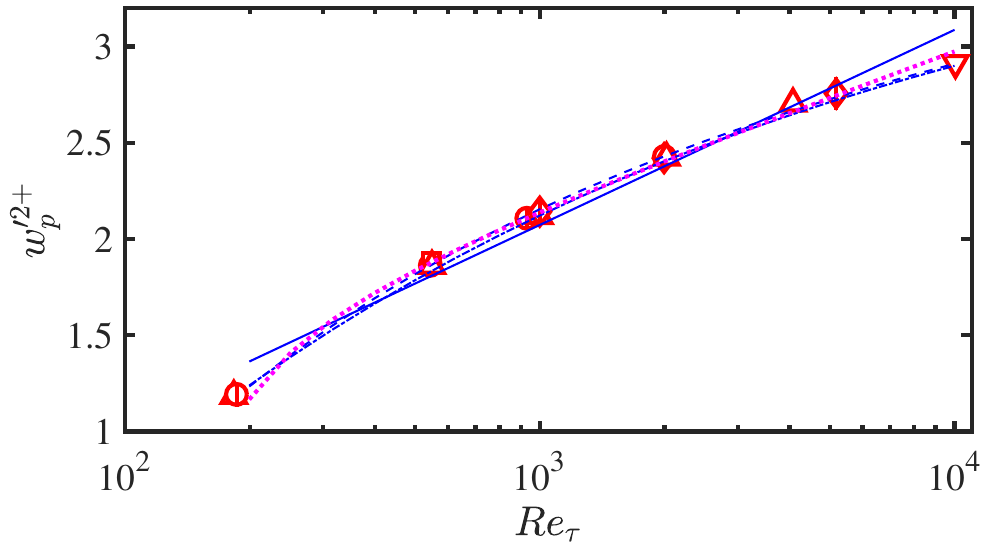}}
\mylab{-.85\columnwidth}{.38\columnwidth}{(d)}%
}%
\caption{%
(a) Intensity of the streamwise velocity fluctuations near the wall. The symbols
mark the position of the near-wall peak, as in table \ref{tab:data}. 
Numerical channels \cite{juanc03,lee:moser:15,hoyas22}. The Reynolds number increases from blue to
red in $\retau=180-10^4$.
(b) As in (a), for the spanwise velocity.
(c) Peak intensity of the streamwise  velocity fluctuations near the wall as a function of the Reynolds number. 
Symbols as in table \ref{tab:data}, with standard deviations as vertical bars when available.
(d) As in (c), for the spanwise velocity.
Lines are different data fits: 
\solid, Logarithmic approximation \cite{tow:61}, $u'^{2+}_p= A_L+B_L\log(Re_\tau)$; 
\dashed, power defect law \cite{chen:sree:21}, $u'^{2+}_p= A_{dP}+B_{dP}Re_\tau^{-1/4}$; 
\chndot, logarithmic defect law \cite{hwang24},  $u'^{2+}_p= A_{dL}+B_{dL}/\log(Re_\tau)$; 
\dotted, shifted logarithm, $u'^{2+}_p= A_S+B_S\log(Re_\tau-150)$.
Coefficients are collected in table \ref{tab:fits}.
}
\la{fig:uupeak_mag}
\end{figure}

%
\begin{table}
\caption{\label{tab:fits}%
Coefficients for the line fits in Fig. \ref{fig:uupeak_mag}.}
\begin{ruledtabular}
\begin{tabular}{lcccc}
& $A_L, B_L$ & $A_{dP}, B_{dP}$  & $A_{dL}, B_{dL}$  &  $A_S, B_S$  \\
\hline
$u'^2$ & 3.75,  0.63 &  11.5, -19.3 & 13.8, -40.0& 4.17,  0.57\\ 
$w'^2$ & -0.97, 0.44 &  3.9, -10.0 & 5.18, -20.9& -0.18,  0.34\\ 
\end{tabular}
\end{ruledtabular}
\end{table}

The fluctuation profiles and peak intensity of the two wall-parallel velocities are given in Fig.
\ref{fig:uupeak_mag}, which shows that they agree reasonably well among different data sets at
similar Reynolds numbers. The growing trend of the profiles is clear in Figs.
\ref{fig:uupeak_mag}(a,b), as is the trend of the peak position to slowly move away from the wall
when the Reynolds number increases (for which there is also some experimental evidence
\cite{Willert:17}).

The peak intensities are collected in Figs. \ref{fig:uupeak_mag}(c,d), with line fits from the
models discussed above, with coefficients either taken from the original publications or fitted
numerically to the data. They separate in two groups. The straight line is the basic logarithm
already discussed above for the attached-eddy model \cite{hoyas06,jim18}, which predicts an infinite
intensity as $\retau\to \infty$. The second group includes the dashed and chaindotted curves in each
figure \cite{chen:sree:21,hwang24}, both of which assume a finite limit for the intensity, as in
Eqs. \r{eq:logdef} and \r{eq:powdef}. The defect formulations appear to represent the data better
than the straight line of the logarithm, especially if we disregard the lowest Reynolds number,
$\retau\approx 180$, but the most striking observation is how similar to each other they are.
Finally, the red dotted lines in the two figures are a shifted logarithm with a virtual origin for
$\retau$, which predicts an unbounded peak at large $\retau$. It is not intended as a serious
proposal, and I am not aware of any theoretical basis for it (although neither is it absurd to shift
the Reynolds number by a transition threshold), but it shows that there are simple approximations
that fit the available data as well as the defect laws, while predicting an infinite limiting value
for the near-wall peak. It emphasizes that simple curve fitting cannot decide the issue.

It may be relevant at this point to estimate which would be the Reynolds number required to
distinguish between the different fits in Figs. \ref{fig:uupeak_mag}(c,d). An order of magnitude
could be how far the logarithmic straight line has to be extended before it reaches the asymptotic
value of any of the two defect laws. The details depend on the approximation and on the variable
chosen, but it is in all cases of the order of $\retau\approx 10^5$. This is at least 15--20 years
in the future for numerical simulations, but it is worth remarking that experiments in pipes
\cite{Willert:17} up to $\retau \approx 4\times 10^4$ have proved inconclusive for this purpose
\cite{hwang24}, and that data from the atmospheric surface layer at $\retau=O(10^6)$, although not
strictly a channel, appear to follow the logarithmic trend reasonably well \cite{met:klew:01}.

\section{The attached-eddy model}\la{sec:AE}
    
The key theoretical contribution to understanding the velocity fluctuations was made by
\textcite{tow:61}, who noted that the usual argument that there is an overlap layer in which neither the viscous
unit of length nor the flow thickness are relevant \cite{millikan38} should also apply to
them. Although the resulting attached-eddy model has been reviewed often, we will recall it
here to identify where the logarithmic prediction could go wrong.
 
The naive argument is that, if there is no scale for lengths, but $u_\tau$ is a scale for the
velocities, the functional dependence of any variable with dimensions of velocity on a variable with
dimensions of length should be logarithmic (see appendix A in Ref. \onlinecite{jim18}). We could
thus expect that $u'/u_\tau \sim \log (y/h)$, which, when particularized at a fixed inner viscous
limit, $y_p^+$, results in $u'_p/u_\tau \sim \log Re_\tau$.
 
There is substantial experimental and numerical evidence for both approximate logarithmic
behaviors \cite{mar:etal:JFMR13}, which, as mentioned above, extend to the fluctuations of the
pressure and of the spanwise velocity\cite{jimhoy08} but, in the absence of a rigorous theory, it is
always possible that higher Reynolds numbers than those currently available may lead to something
different. 

Moreover, there are logical flaws in the previous argument. Most obviously, an argument similar to
the one used for $u'$ applies to any power of the velocity, and some selection rule is needed to
decide which power to use. In addition, some reason needs to be found for why the logarithmic
behavior applies to $u'$ and $w'$ but, as mentioned above, not to $v'$, and a justification is
required for why $\utau$ is the right scaling unit for the velocity fluctuations. 
 
In general, it is unwise to use similarity arguments without a dynamical model, and
\textcite{tow:61} proposed that the logarithm is implemented by the superposition of a self-similar
family of `attached' Reynolds-stress-carrying eddies linking the wall to the interior of the flow.
In the absence of a fixed length scale, their height is proportional to their wall-parallel size
($\lambda_x$ or $\lambda_z$), while their intensity is $O(u_\tau)$ because each eddy family is
responsible for carrying the tangential Reynolds stress $(-uv)$ at one distance from the wall. This
is supported by observations: the solid contours in Figs. \ref{fig:cospec}(a,b) are the
premultiplied cospectra, $\Phi_{uv}^+(k_x)$ and $\Phi_{uv}^+(k_z)$ of the tangential Reynolds
stress, respectively drawn as functions of $y$ and of the streamwise or spanwise wavelengths. The
contours are normalized with the friction velocity, and the figures show that a stress $-uv\sim
O(u_\tau^2)$ is concentrated along a spectral ridge in which $\lambda_x\approx 25 y$ and
$\lambda_z\approx 5 y$, and which extends from a minimum wavelength that scales in wall units to an
outer one that scales with $h$. Since it can be shown that the streamwise and wall-normal velocities
are well correlated in the regions of the $\lambda-y$ plane where the Reynolds stresses are
substantial, this extends the role of $\utau$ from a velocity scale for the overall intensity to one
for individual spectral bands, although the argument may not apply to 'sterile' scales without
tangential stress.

\begin{figure}
\centerline{%
\raisebox{0mm}{\includegraphics[width=.7\columnwidth,clip]%
{\figpath 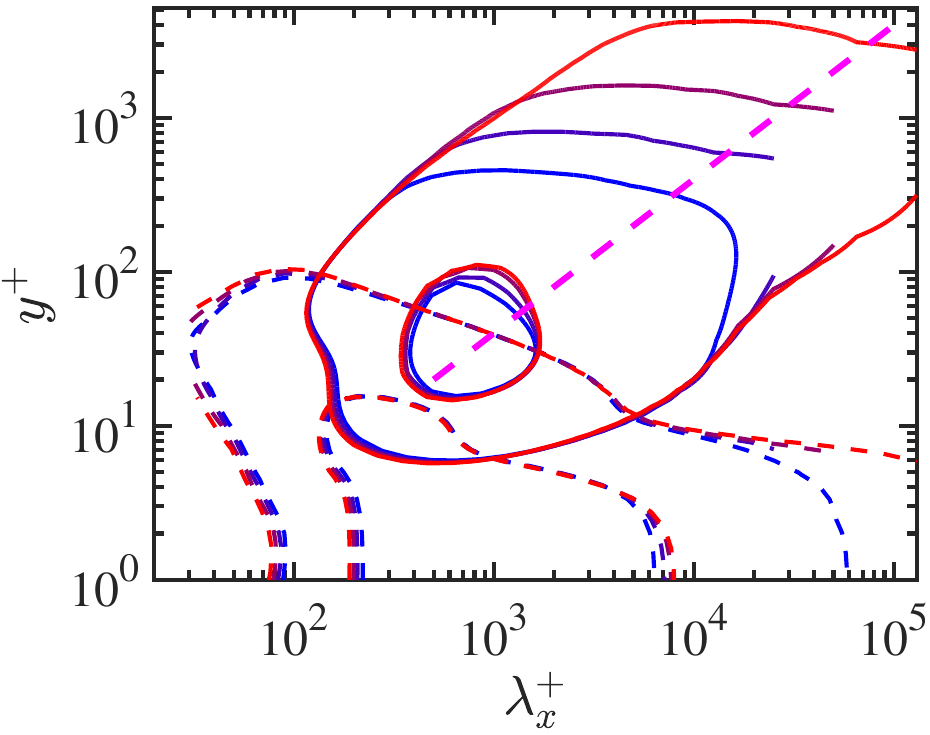}}
\mylab{-.55\columnwidth}{.47\columnwidth}{(a)}%
}%
\vspace{2mm}%
\centerline{%
\raisebox{0mm}{\includegraphics[width=.7\columnwidth,clip]%
{\figpath 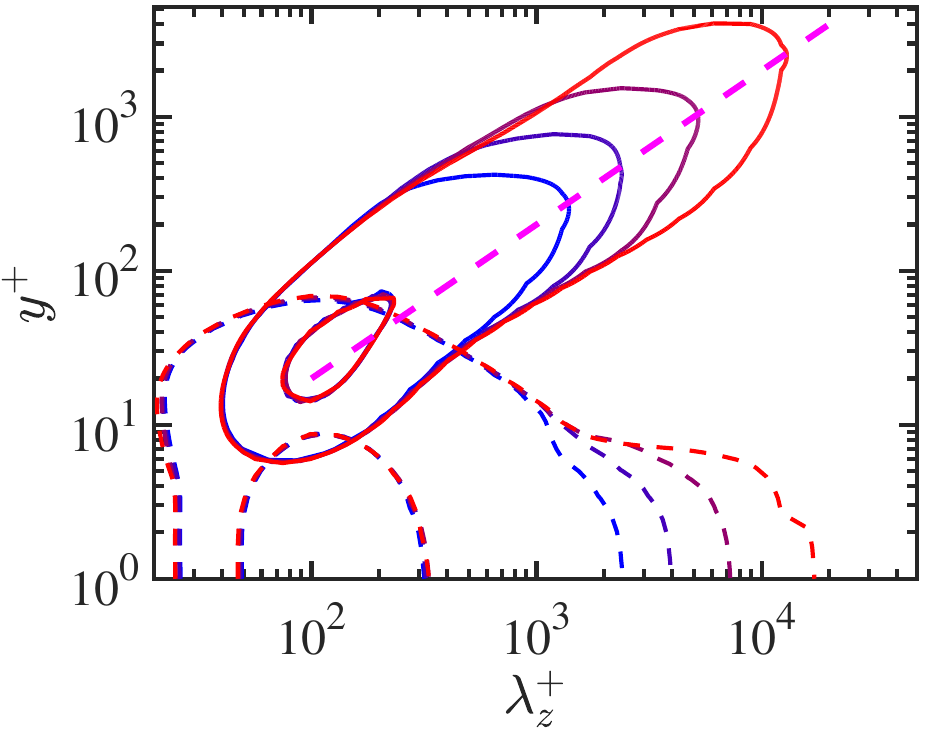}}
\mylab{-.55\columnwidth}{.47\columnwidth}{(b)}%
}%
\vspace{2mm}%
\centerline{%
\raisebox{0mm}{\includegraphics[width=.7\columnwidth,clip]%
{\figpath 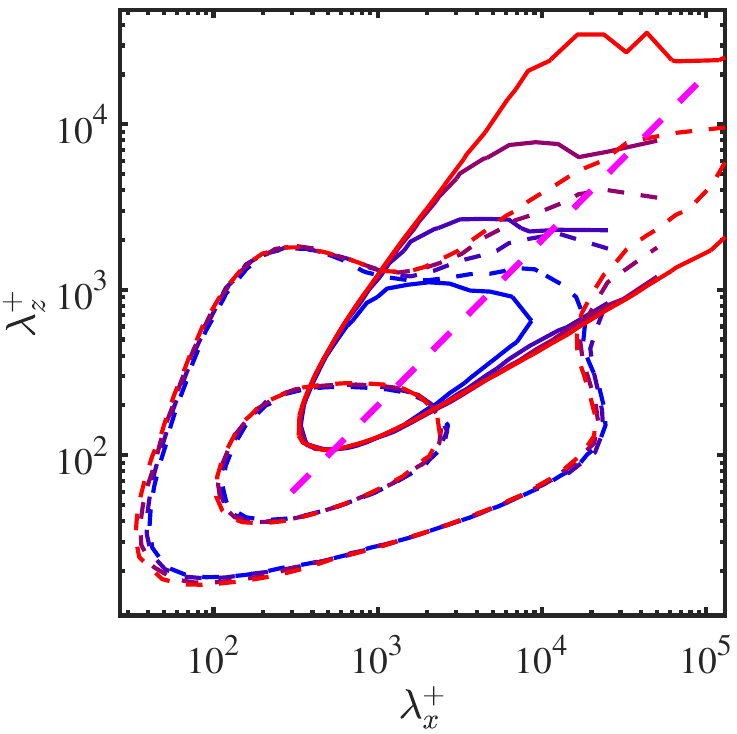}}
\mylab{-.55\columnwidth}{.61\columnwidth}{(c)}%
}%
\caption{%
(a) Solid contours are premultiplied cospectra of the tangential Reynolds stress, 
$-\Phi_{uv}^+(\lambda_x,y)=[0.2,\,0.8]$ times the common maximum of the four cospectra.
Dashed ones are premultiplied spectra of the spanwise vorticity,
$\Phi_{\omega_z\omega_z}^+(\lambda_x,y)=[0.05,\, 0.4]$ the common maximum. 
The dashed diagonal is $\lambda_x=25y$.
(b) As in (a), but as functions of $\lambda_z$. The dashed diagonal is $\lambda_z=5y$.
(c) Two-dimensional integrated spectra as functions of $(\lambda_x,\lambda_z)$: 
Solid contours are $-\int \Phi_{uv}^+\dd y^+$ integrated over $y\in [0.005,\,0.5]\lambda_x$. Contours
are $[0.2,\,0.8]$ the common maximum.
Dashed contours are premultiplied spectra of the spanwise vorticity, $\int \Phi_{\omega_z\omega_z}^+ \dd
y^+$, integrated over $y^+\in [0,\,20]$. Contours are $[0.05,\,0.4]$ the common maximum.
The dashed diagonal is $\lambda_x=5 \lambda_z$.
Numerical channels \cite{lozano14,lee:moser:15}, with the Reynolds number increasing from blue to
red: $Re_\tau=550,\,1000,\,2000,\, 5200$. }
\la{fig:cospec}
\end{figure}

Because there is no fixed length scale, these bands are logarithmic (e.g. from $\lambda_x$ to
$2\lambda_x$), and the number of eddy families found at a given distance $y$ from the wall is the
number of logarithmic bands required to cover the range of lengths from $y$ to the flow thickness,
$O(h)$, which increases logarithmically with $h/y$. If we further assume that the velocity
fluctuations of the different bands are uncorrelated, their variances add, resulting in a
logarithmic behavior with the Reynolds number for $u'^{2+}(y)\sim \log(y/h)$ and for $u'^{2+}(y_p)$.
\textcite{tow:61} distinguishes between `inactive' attached variables, such as $u'$ and $w'$, which
are not sufficient to generate tangential Reynolds stress and are only damped by the wall in a thin
viscous layer, and `active' ones, like $v'$, that contribute to the stress but are inhibited by
impermeability at distances from the wall comparable to their size. Only the former should have
logarithmic profiles.

The difference between the two types of variables can be seen in Figs. \ref{fig:cospec}(a,b), where
the dashed isolines are spectra of the spanwise vorticity. 
\notyet{\textcite{bradshaw67} argued that, in the absence of other terms of the momentum equation,
the flow at long wavelengths, and at distances from the wall below the Reynolds-stress ridge, can only
be driven by the pressure footprint of active eddies further from the wall and should therefore be
essentially irrotational. It is indeed clear from \ref{fig:cospec}(a,b) that wavelengths longer than
$\lambda_x^+\approx 10^4$ are approximately irrotational above $y^+\approx 10$, and Fig. 12(d) in
Ref. \onlinecite{jim18} shows that the pressure is a deep variable, correlated from the wall to the
Reynolds-stress ridge.} But potential flow cannot satisfy the no-slip condition at the wall, and a
viscous rotational layer appears at long wavelengths below $y^+\lesssim 10$. Since $\omega_z\approx
-\p_y u$ in that region, this implies that non-trivial fluctuations of $u$ reach the vicinity of the
wall at those wavelengths.

The solid contours in Fig. \ref{fig:cospec}(c) are two-dimensional Reynolds-stress cospectra
integrated over the active band of wall distances corresponding to each wavelength, and the dashed
ones are spectra of the spanwise vorticity integrated over the viscous near-wall layer. The figure
strongly supports that the latter are the effect of the detached Reynolds stresses, although not
necessarily at the same distance from the wall for all wavelengths. \notyet{The same is implied by
the advection velocity of the long wavelengths near the wall, which is not the local mean flow velocity
\cite{jcadvel,jim18}, but the velocity at $y/h\approx 0.3-0.4$. However, the conclusion that they
are driven by pressure is less clear. Pressure correlations only attach to the wall for
$\lambda_x/h\lesssim 4$, and cannot therefore drive longer wavelengths \cite{sillero14}. In fact, it
will be seen below that those scales are better described as internal turbulent
layers, presumably driven from above by turbulent diffusion \cite{jim18}.}

Families of self-similar attached eddies, as well as the distinction between active and inactive
motions, have been observed and characterised in some detail numerically
\cite{jc06_vor,lozano-Q,jim18} and experimentally \cite{adrmeintom00,tomadr03,adr07,DeshMonMar:21}.

There are several debatable points in these arguments, most of which have been discussed in the
literature and will not be repeated here, but some of which deserve closer attention. A well-known
limitation of the attached-eddy model is that it does not include viscosity, whose effects are
lumped in the rule that something happens below $y^+_p=15$. This has occasionally been suggested
as responsible for scaling failures \cite{hwang24}. We have seen that what viscosity does is to
generate the thin vortex layers in Fig. \ref{fig:cospec}, but how they are maintained, their
dimensions, and their effect are unclear, and we mentioned in the introduction that their
stability is problematic if the fluctuations become too strong.

However, the overall conclusion from the previous discussion remains that the reason why velocity
fluctuations do not scale well is that they contain a wide range of wavelengths, even when they are
very close to the wall. This often-made point \cite{per:cho:82,phc86,mar:kun:03,hoyas06} is clear
from the spectra in Fig. \ref{fig:uu15} of the near-wall streamwise velocity at $y_p^+=15$. There is a
universal `core', centered at $\lambda_x^+\times \lambda_z^+ \approx 1000\times 100$, which does not
reach above $y^+\approx 100$ (Fig. \ref{fig:cospec}) and collapses well across $\retau$. That its
dimensions scale in wall units shows that it is controlled by viscosity
\cite{jmoin,Hamilton95,Waleffe97}, but it is accompanied by larger-scale spectral tails that extend
to $\lambda \sim O(h)$, and which are responsible for the extra energy of the overall fluctuations.
The dynamics of these tails, which does not have to be the same as for the viscous core, has to be
studied if the overall peak intensity is to be understood.

This wide range of scales is a problem for asymptotic models that seek to represent the flow in
terms of an outer `regular' turbulence and a small-scale object near the wall, since the latter does not exist
as such in real flows. In fact, given that the total peak energy includes contributions from a wide
range of wavelengths, it is unclear whether it makes sense to speak of a single velocity and length
scale for it.

\begin{figure}
\centerline{%
\raisebox{0mm}{\includegraphics[width=.75\columnwidth,clip]%
{\figpath 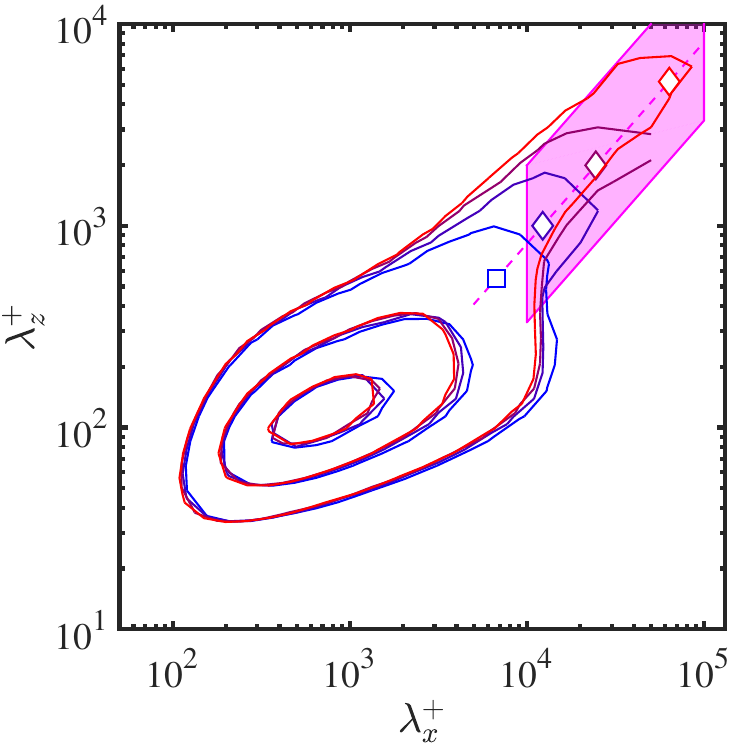}}
\mylab{-.55\columnwidth}{.60\columnwidth}{(a)}%
}%
\caption{%
Premultiplied energy spectrum of the streamwise velocity, $\Phi_{uu}^+$ at $y^+=15$, versus the
wall-parallel wavelengths.
Numerical channels \cite{lozano14,lee:moser:15}, with the Reynolds number increasing from blue to
red: $Re_\tau=550,\,1000,\,2000,\, 5200$. Contours are $[0.15,\, 0.4,\,0.8]$ times the common maximum of
the four spectra. The dashed diagonal is $\lambda_x=12\lambda_z$, and symbols are $\lambda_z=h$.
The translucent patch is used in Fig. \ref{fig:specden}.
}
\la{fig:uu15}
\end{figure}

\section{The large scales}\la{sec:results}

\begin{figure}
\centerline{%
\raisebox{0mm}{\includegraphics[width=.73\columnwidth,clip]%
{\figpath 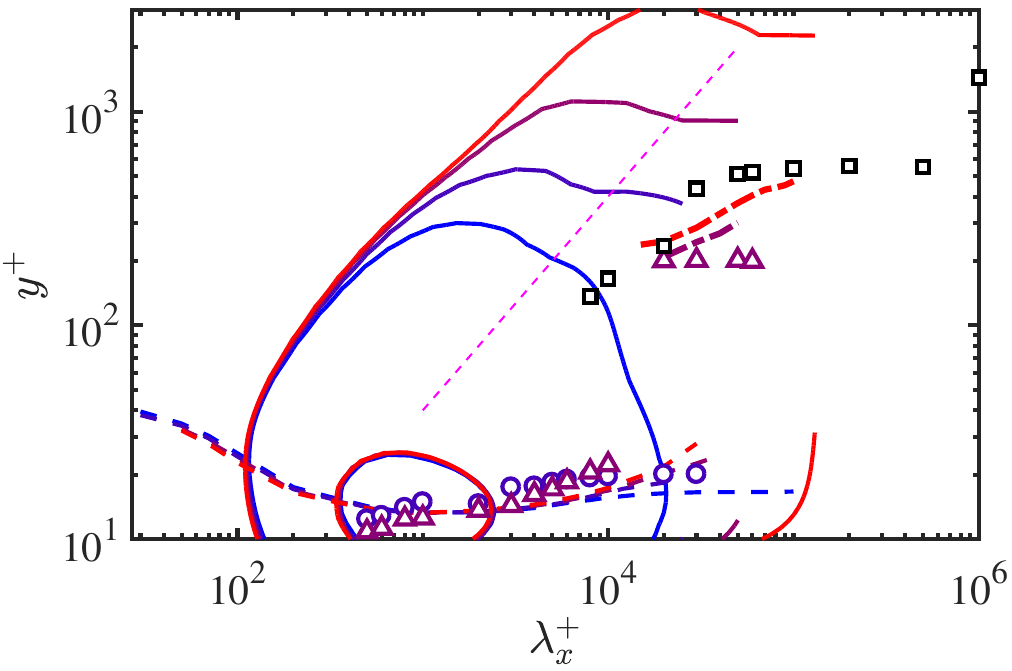}}
\mylab{-.76\columnwidth}{.43\columnwidth}{(a)}%
\hspace*{0mm}%
}%
\vspace{0mm}%
\centerline{%
\raisebox{0mm}{\includegraphics[width=.73\columnwidth,clip]%
{\figpath 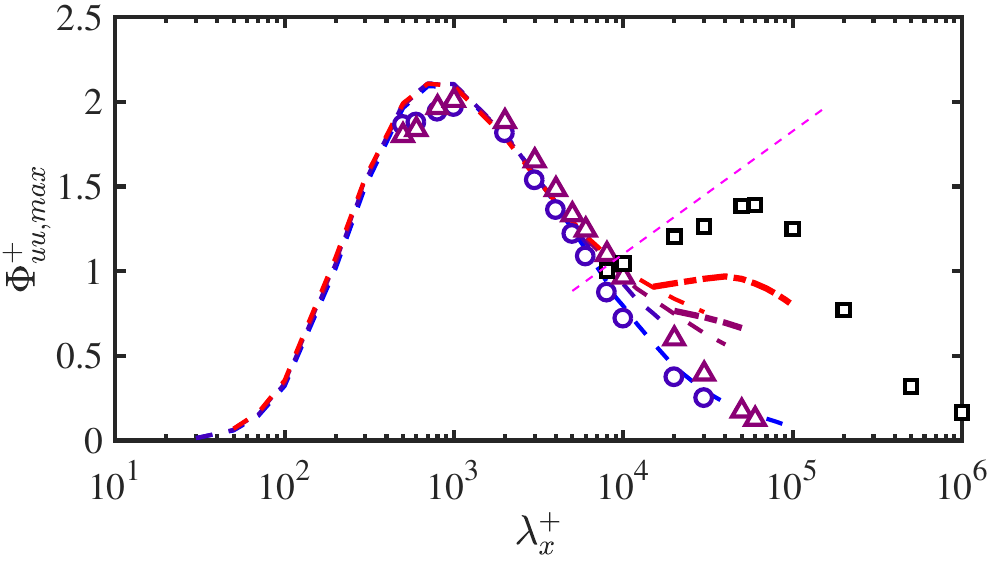}}
\mylab{-.77\columnwidth}{.37\columnwidth}{(b)}%
}%
\vspace{0mm}%
\centerline{%
\raisebox{0mm}{\includegraphics[width=.73\columnwidth,clip]%
{\figpath 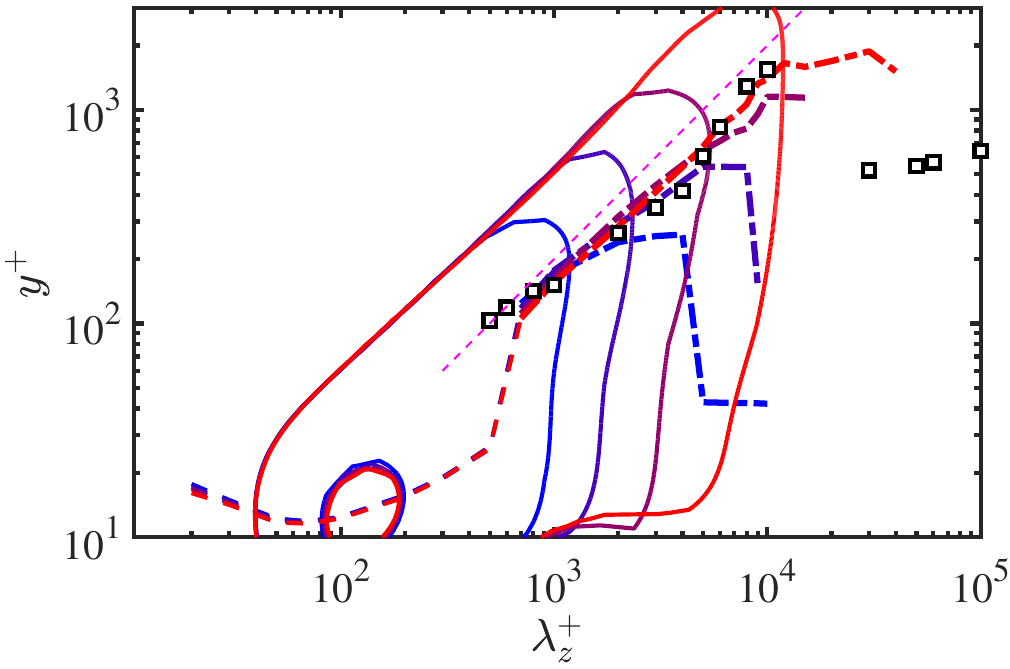}}
\mylab{-.76\columnwidth}{.43\columnwidth}{(c)}%
\hspace*{0mm}%
}%
\vspace{0mm}%
\centerline{%
\raisebox{0mm}{\includegraphics[width=.72\columnwidth,clip]%
{\figpath 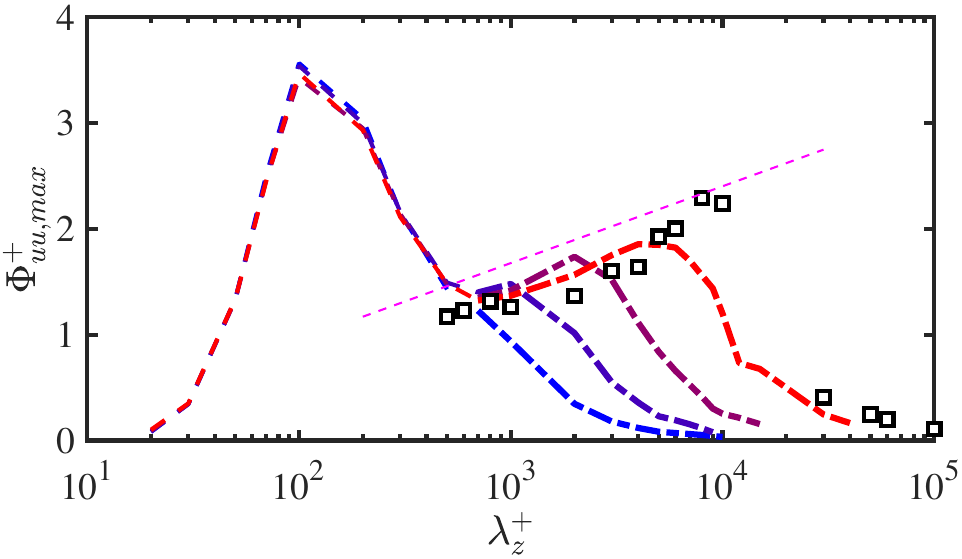}}
\mylab{-.77\columnwidth}{.37\columnwidth}{(d)}%
}%


%
\caption{%
(a,c) Contours are the premultiplied spectrum of the streamwise velocity. The contours are $[0.2,\,
0.8]$ times the overall maximum of all the $\Phi_{uu}^+$.
Dashed and chaindotted lines are the location of the inner and outer maxima of $\Phi_{uu}$ at each
wavelength. Numerical channels \cite{lozano14,lee:moser:15}, with the Reynolds number increasing
from blue to red: $Re_\tau=550,\,1000,\,2000,\, 5200$.
Symbols are boundary layers from Ref. \onlinecite{kunkel03}: \opencirc, $Re_\tau=1014$; \opentriangle,
1844; or from Ref. \onlinecite{DeshMonMar:21}, \opensquare, $Re_\tau=14000$.
(b,d) Maximum  of $\Phi_{uu}^+$ versus wavelength, with the same lines and symbols as above.
(a,b) Against $\lambda_x$. (c,d) Against $\lambda_z$.
The dashed magenta line in (a) and (c) are 
$\lambda_x^+=25y^+$ and $\lambda_z^+=5y^+$, respectively. The ones in (b) and (d) are
$\Phi_{uu,max}^+ = 0.315\log (\lambda_x^+) -1.8$ and $\Phi_{uu,max}^+ = 0.315\log (\lambda_z^+)
-0.5$.
}
\label{fig:spec1d}
\end{figure}

Figures \ref{fig:spec1d}(a,c) are similar to the cospectra in Figs. \ref{fig:cospec}(a,b), but drawn
for the streamwise velocity component. To facilitate comparison, the dashed diagonals in the two
figures are the same, and show that the upper limit of the kinetic energy and of the Reynolds stress
are similar, but it is clear from Figs. \ref{fig:cospec} and \ref{fig:spec1d} that the streamwise
velocity spectrum extends to the wall, while the cospectrum stays away from it. \notyet{Notice, in
particular, the different behavior of the larger wavelengths.} 

Figure \ref{fig:spec1d} includes as thick dashed lines and symbols the wall-normal location of the
maximum of $\Phi_{uu}$ at each wavelength. We saw when discussing Fig. \ref{fig:uupeak_mag} that the
energy peak drifts slowly away from the wall with the Reynolds number, and Figs.
\ref{fig:spec1d}(a,c) show a clearer dependence on the wavelengths. The symbols are from
experimental boundary layers \cite{kunkel03,DeshMonMar:21} and, although sparser near the wall, are
compatible with the numerics. The inner maximum at the shorter wavelengths marks the edge of the
viscous layer, and is superseded beyond $\lambda_x^+\approx 10^4$ by an outer maximum located at
$y_{max}\approx \lambda_z/5$ or $\lambda_x/25$ (Fig. \ref{fig:spec1d}a,c). Note that this implies
that the transition to an outer peak always takes place when $y_{max}^+\approx 200$, which can be
interpreted as the thickness at which the near-wall viscous layers become unstable.

The amplitude of these maxima is shown in Figs. \Ref{fig:spec1d}(b,d). It collapses well with
$\retau$ in the viscosity-dominated region, $\lambda_x^+\lesssim 10^4$ or $\lambda_z^+\lesssim
10^3$, but not at the larger wavelengths associated with the outer maximum. Their growth with
$\lambda$ can be interpreted as a wavelength-by-wavelength counterpart to the logarithmic growth of
$u'^2_p$ with $\retau$, represented here by the range of wavelengths over which the one-dimensional
spectrum is summed. We have added to Figs. \ref{fig:spec1d}(b,d) logarithmic approximations to this
growth, using half the slope in Fig. \ref{fig:uupeak_mag} for $u'^2_p$, on the assumption that half
of the peak growth is due to the wider range of $\lambda_x$ and the other half to the wider range of
$\lambda_z$. These logarithms match the data relatively well, but they should only be considered as
aids to the eye, subject to the same caveats as the curve fits in Fig. \ref{fig:uupeak_mag}(a,b).

Fortunately, more can be said about the large flow scales. The outer maximum of the longest
structures in Fig. \ref{fig:spec1d}(a) represents layers of dimensions $(\lambda_x\times
y\times\lambda_z)^+= O(20000\times 500\times 4000)$, which are thin both with respect to the channel
height $(\retau=5200)$ and to their own length. They are also deep enough to be fully turbulent.
Internal turbulent layers are common when wall-bounded flows cross boundaries between different
types of wall or are otherwise perturbed, and have been extensively studied in meteorology
\cite{Garr90,BouEtal2020} and for heterogeneous surfaces \cite{LiEtal:22}. Using a rough
approximation \cite{BouEtal2020}, there is an internal equilibrium layer (EL) whose thickness after
time $\Delta t$ grows to $\delta_{EL}\approx \utau \Delta t/4$, and satisfies the universal velocity
profile corresponding to the friction velocity that develops after the perturbation. Since the
lifetime of the stress-carrying structures is \cite{lozano-time} $\Delta t\approx \Delta y/\utau
\approx \lambda_z/\utau$, the thickness of the EL generated by perturbations of width $\lambda_z$ is
$\delta_{EL}\approx \lambda_z/4$, which approximately agrees with the location of the outer maximum
in Fig. \ref{fig:spec1d}(c) .

\begin{figure}
\centerline{%
\raisebox{0mm}{\includegraphics[width=.68\columnwidth,clip]%
{\figpath 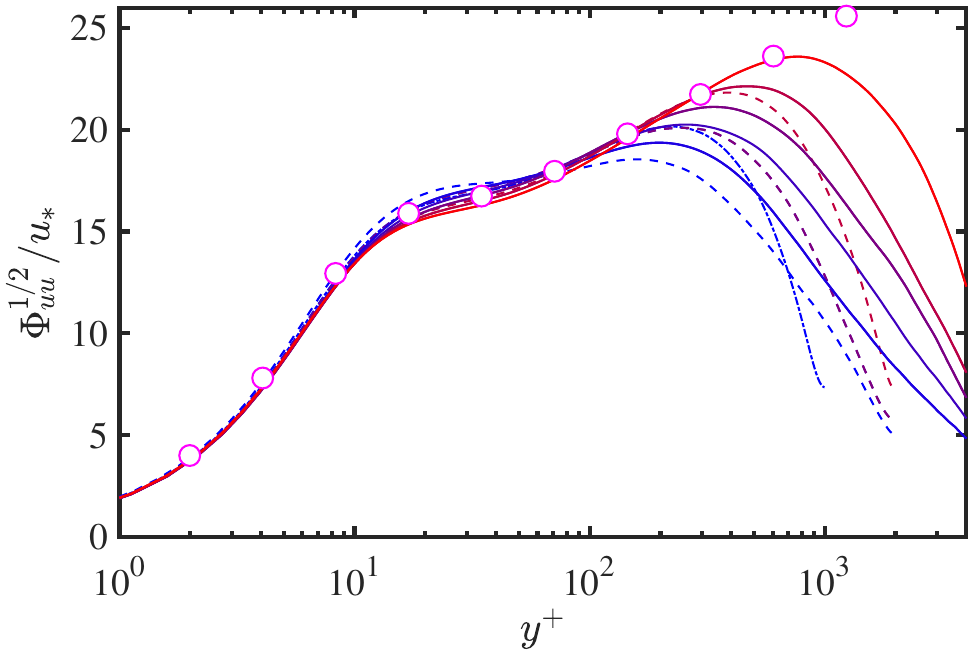}}
\mylab{-.77\columnwidth}{.38\columnwidth}{(a)}%
}%
\vspace{1mm}%
\centerline{%
\raisebox{0mm}{\includegraphics[width=.70\columnwidth,clip]%
{\figpath 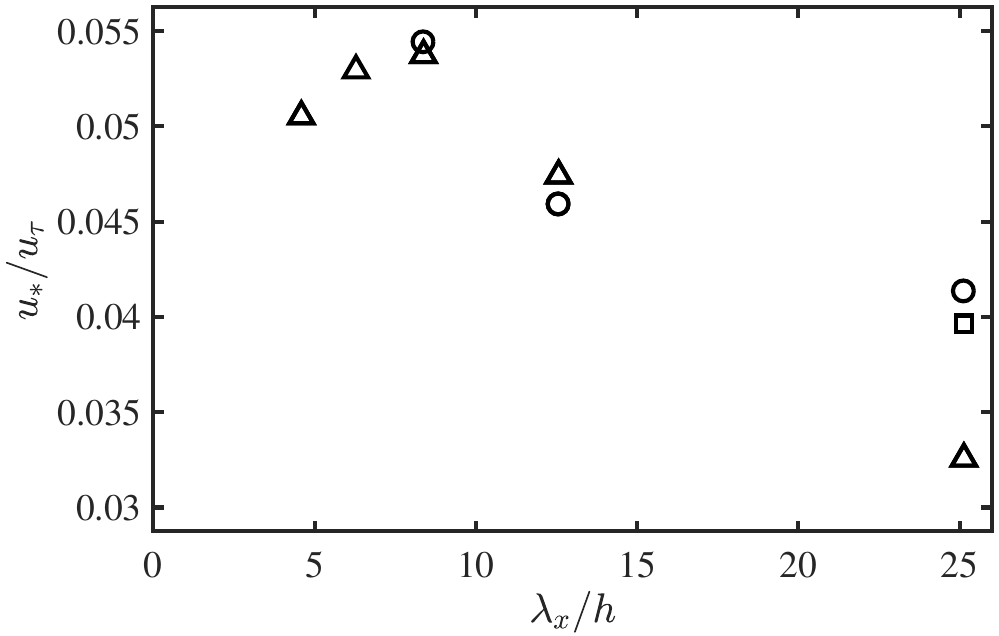}}
\mylab{-.77\columnwidth}{.35\columnwidth}{(b)}%
}
\vspace{1mm}%
\centerline{%
\raisebox{0mm}{\includegraphics[width=.70\columnwidth,clip]%
{\figpath 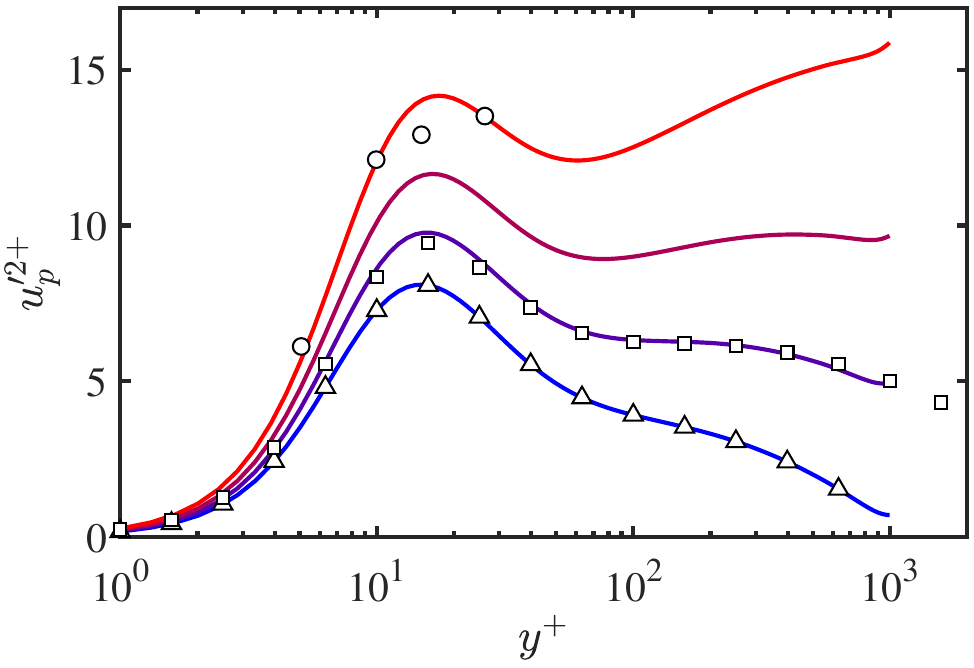}}
\mylab{-.77\columnwidth}{.35\columnwidth}{(c)}%
}%
\caption{%
(a) Spectral density of the streamwise velocity for various wavelengths, from $\lambda_x^+=2\times
10^4$ to $\lambda_x=25h$, from blue to red, integrated in $\lambda_z$ over the diagonal band in Fig
\ref{fig:uu15}, defined as $\lambda_z \in\lambda_x/[30,\, 5]$. The symbols are $G$ in Eq.
\r{eq:allpertur}, computed from the mean velocity profile in \cite{lee:moser:15}.
(b) Friction velocity perturbation intensities obtained from fitting (a). 
%
Channels
\cite{lee:moser:15}: (\chndot\ and \opensqr) $\retau=1000$; (\dashed\ and \opencirc) $\retau=2000$;
(\solid\ and \opentriangle) $\retau=5200$.
(c) Lines are fluctuation profiles estimated from Eq. \r{eq:totuu}. From bottom to top:
$\retau=1000,\,10^4,\, 10^5$ and $2\times 10^6$. Symbols are channel simulations
\cite{lee:moser:15,hoyas22} and the atmospheric surface layer \cite{metetal01} at those Reynolds
numbers. The dashed line is $u'^{2+}_s $, estimated from $\retau=1000$.
}
\label{fig:specden}
\end{figure}

The model is that large-scale fluctuations with wavelengths of the order of $\lambda^+_x\approx
5\lambda^+_z\gtrsim 10^4$ (Fig. \ref{fig:cospec}c) are equilibrium turbulent boundary layers
satisfying the universal profile $U=u_\tau F(y^+)$ with a perturbed friction velocity, and that, if
they are viewed as perturbations to the overall mean velocity, they can be modeled as weak
perturbations of $\utau\to \utau+\delta \utau$. Upon linearization,
\beq
(\delta U)= (\delta \utau) (F+y^+\dr F/\dr y^+)
\equiv (\delta \utau) G( y^+),
\la{eq:allpertur}
\eeq
which links the rms intensity of the velocity fluctuations to the rms perturbation of the friction
velocity, $(\delta\utau)'=u_*$.  
In the viscous layer, $F=y^+ $, and
\beq
(\delta U)'/u_*= 2y^+. 
\la{eq:buffpertur}
\eeq
This is essentially the modulation described in \cite{Mar_Sci10}, which was shown in
\cite{jim12_arfm} to reduce in the buffer layer to a modulation of $\utau$. We extend it here to the
logarithmic layer, where $F$ satisfies \r{eq:loglaw}, and
\beq
(\delta U)'/u_* = 
\kappa^{-1}\left(1+A\kappa+\log y^+\right).
\la{eq:logpertur}
\eeq
These equations are tested in figure \ref{fig:specden}(a) for spectral densities within the patch of
large-scale tails in Fig. \ref{fig:uu15}. The perturbation intensity, $u_*$, of the friction
velocity has been adjusted for each wavelength to fit Eq. \r{eq:allpertur} to the spectral profile
from the wall to the location of the spectral maximum, but the definition of $F$ has not been
modified. The symbols in Fig. \ref{fig:specden}(a) are $G(y^+)$ in Eq. \r{eq:allpertur}, computed
from the mean velocity $F(y^+)$ of a turbulent channel, and the agreement is excellent, supporting
the assumption of turbulent equilibrium internal layers. Note that the lowest $\retau=550$ is not in
the figure because it never develops an outer peak. This model was probably first used by
\textcite{bradshaw67} to explain the differences between boundary layers subject to different
streamwise pressure gradients. Some time later, \textcite{spalart88} tested it on simulations of a
zero-pressure-gradient boundary layer at $\retau\approx 400$, and concluded that its large scales
were better represented by laminar oscillating Stokes layers, but the comparison with Fig.
\ref{fig:spec1d} shows that his Reynolds number was too low, and his wavelength too short
$(\lambda_x^+\approx 10^3)$, falling in the viscous range of Fig. \ref{fig:spec1d}(a).

More recently, \textcite{piroz:24} has applied the equilibrium layer model to pipes, with results
similar to Fig. \ref{fig:specden}. However, the amplitude of his perturbations depends on
$\lambda^+$, while Fig. \ref{fig:specden}(b) shows that $u_*$ scales with $\lambda_x/h$ rather than
in wall units. The reason for this discrepancy is unknown, and deserves further investigation. The
rms of the fluctuations of $\utau$ tend to some non-zero value when $\lambda_x\to 0$, so that the
fluctuation velocity profile can be approximated as
\beq
u'^{2+} (y^+) = u'^{2+}_s (y^+) + 
      G^{2+}(y^+) \int_{\lambda_1/h}^{\lambda_2/h} u_*^{2+}(\lambda/h)  \dd\log \lambda, 
\la{eq:totuu}
\eeq
where $ G(y^+)$ is defined in Eq. \r{eq:allpertur}, $\lambda_1=\max(\lambda_{in}, 25y)$, and
$u'^{2+}_s$ is the contribution of scales whose $\lambda_x$ is shorter than the wavelength,
$\lambda_{in}$, at which the viscous sublayer becomes unstable and fluctuations have to be modeled
as turbulent profiles. It is important to note that the lower limit of this integral scales in wall
units ($\lambda_{in}^+\approx 2\times 10^4$ in Fig. \ref{fig:spec1d}), while the upper one either extends to
infinity, or scales in outer units ($\lambda_2=25h$ in our case, because of the
length of our computational box). Assuming that the small-scale contribution to Eq. \r{eq:totuu} is
independent of $\retau$, and that the integral of $u_*^2$ stays bounded at $\lambda\to\infty$, the
dominant contribution to Eq. \r{eq:totuu} at high $\retau$ comes from the lower limit of the
integral,
\beq
u'^{2+} (y^+) \approx u'^{2+}_s (y^+) + 
      G^{2+}(y^+) u_*^{2+}(0) \log(\retau/\lambda_{in}^+). 
\la{eq:totuup}
\eeq
It is important to realize that $U+\delta U$ in Eq. \r{eq:allpertur} is the mean profile of a
perturbed boundary layer. These faster- or slower-than-average local equilibrium layers have their
own small-scale perturbations that are modulated by the outer Reynolds-stress structures, but those
are second-order effects, negligible with respect to the mean profile. The maximum fluctuation
intensity at the inner energy peak is $u'^{2+}_p\approx 9$ in our data base, while the mean velocity at
that point is $U^{2+}(y_p) \approx 100$. The fluctuation profiles are only relevant if the mean
profile can be considered fixed, but any perturbation of the latter overwhelms the modulation of the
small scales.

Eq. \r{eq:totuup} is also a logarithm that diverges as $\retau\to\infty$, but its most interesting
aspect is that the form of the large-scale correction is not a peak near the wall, but something
similar to the mean velocity profile of a regular boundary layer, so that the fluctuation peak will
be absorbed into something closer to a plateau at high $\retau$. Although estimating this behavior
necessarily implies extrapolation from lower Reynolds numbers, Fig. \ref{fig:specden}(c) plots Eq.
\r{eq:totuu} for several large $\retau$. The lowest curve in the figure, $\retau=1000$ is used to
estimate the viscous contribution, $u'_s$, displayed in the figure as a dashed line, and is
therefore automatically fitted. But the corrections for the rest of the curves are computed by
estimating the integral in Eq. \r{eq:totuu}, using a lineal least-square approximation to $u_*$ in
Fig. \ref{fig:specden}(b). The agreement with our highest Reynolds number, $\retau=10^4$ is
excellent, and it is intriguing that the uppermost curve in the figure, intended to approximate data
from the atmospheric surface layer\cite{metetal01}, also appears to fit well. The fluctuation
profile in this case is already very different from that at lower Reynolds numbers, including a
second outer maximum of $u'$ that requires experimental confirmation. Although the implied
$\retau=O(10^6-10^7)$ are well in the future for laboratory or numerical flows, they are not out of
range for geophysical ones \cite{met:klew:01,metetal01}. In fact, some measurements in the
atmospheric surface layer show a clear double peak in the profile of $u'$, but there is some
uncertainty as to whether this may be partly due to damping of the near-wall peak by roughness,
buoyancy, or instrumental effects \cite{metetal07}.

\begin{figure}
\centerline{%
\raisebox{0mm}{\includegraphics[width=.7\columnwidth,clip]%
{\figpath 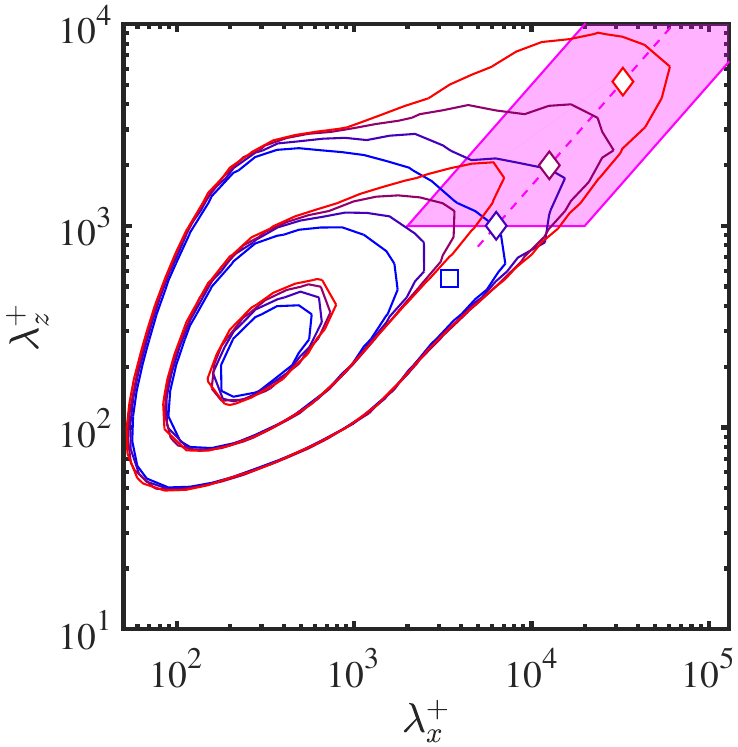}}
}%
\caption{%
Premultiplied energy spectrum of the spanwise velocity, $\Phi_{ww}^+$ at $y^+=50$, versus the
wall-parallel wavelengths.
Numerical channels \cite{lozano14,lee:moser:15} with the Reynolds number increasing from blue to
red: $Re_\tau=550,\,1000,\,2000,\, 5200$. Contours are $[0.15,\, 0.4,\,0.8]$ of the common maximum of
the four spectra. The dashed diagonal is $\lambda_x=6\lambda_z$ and symbols are $\lambda_z=h$.
The translucent patch is used in Fig. \ref{fig:spec1dw}(c,d).
}
\la{fig:ww15}
\end{figure}

\section{The spanwise velocity}\la{sec:w}

Up to now we have mostly dealt with the streamwise velocity fluctuations, but it is clear from Fig.
\ref{fig:uupeak_mag} that the spanwise velocity also has a potentially infinite limit. Figure
\ref{fig:ww15} displays the two-dimensional spectrum of $w'$, and shows that the reason is also the
effect of large structures. Their maximum size also scales with $h$ rather than in wall units, but
they are wide rather than long, as required by continuity \cite{bat53}.

\begin{figure}
\centerline{%
\raisebox{0mm}{\includegraphics[width=.73\columnwidth,clip]%
{\figpath 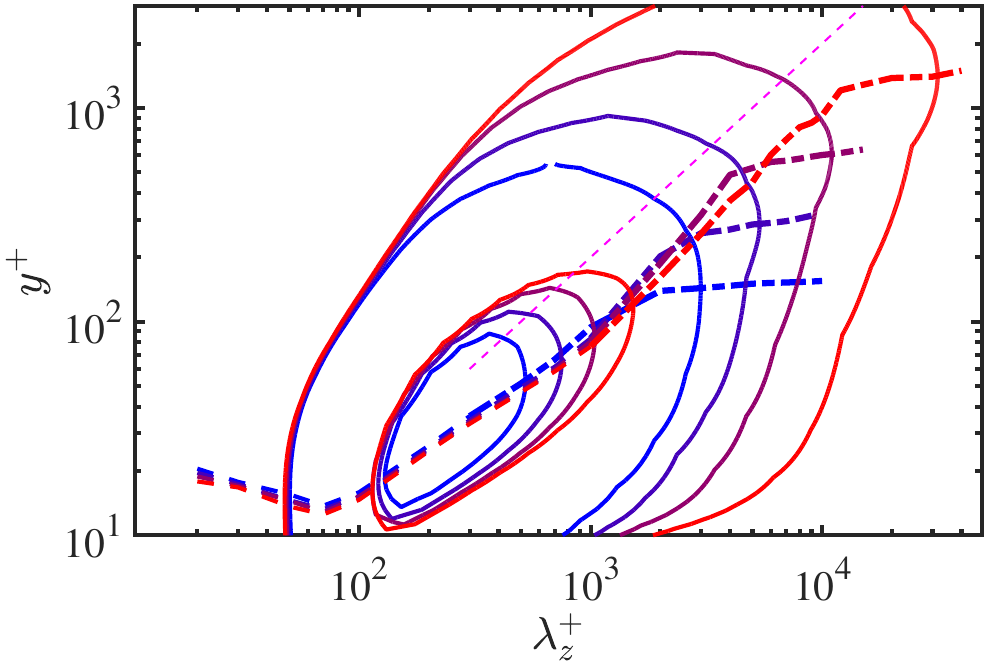}}
\mylab{-.76\columnwidth}{.43\columnwidth}{(a)}%
\hspace*{1mm}%
}%
\vspace{1mm}%
\centerline{%
\raisebox{1mm}{\includegraphics[width=.73\columnwidth,clip]%
{\figpath 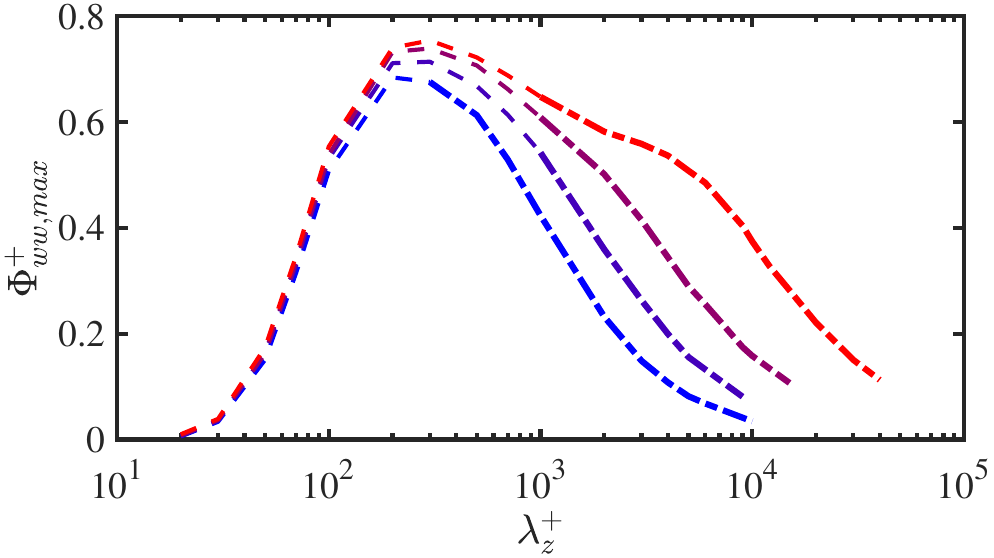}}
\mylab{-.77\columnwidth}{.37\columnwidth}{(b)}%
}%
\vspace{1mm}%
\centerline{%
\raisebox{0mm}{\includegraphics[width=.70\columnwidth,clip]%
{\figpath 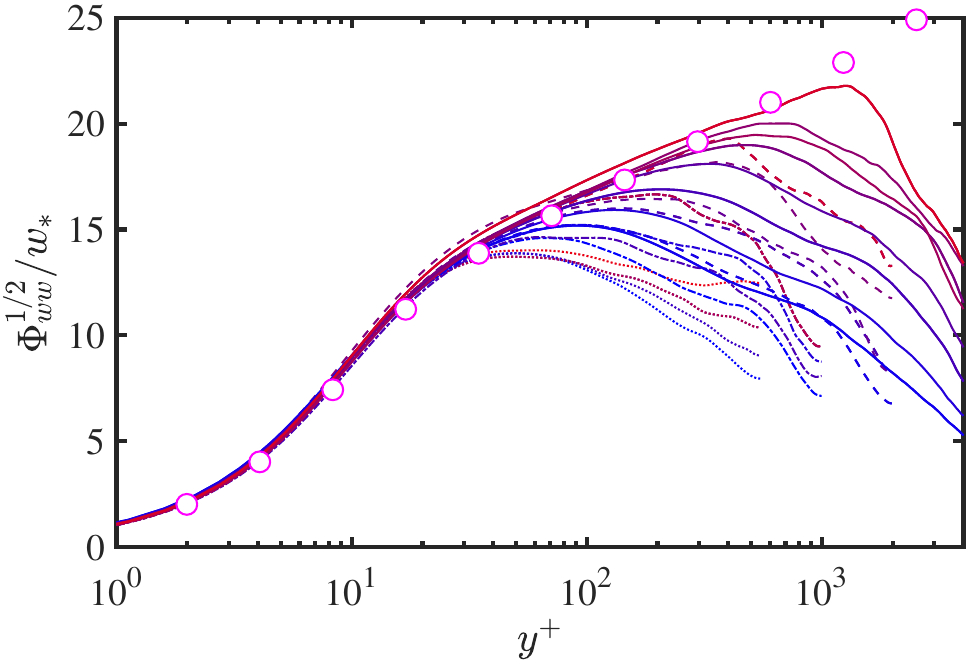}}
\mylab{-.77\columnwidth}{.38\columnwidth}{(c)}%
}%
\vspace{1mm}%
\centerline{%
\raisebox{0mm}{\includegraphics[width=.73\columnwidth,clip]%
{\figpath 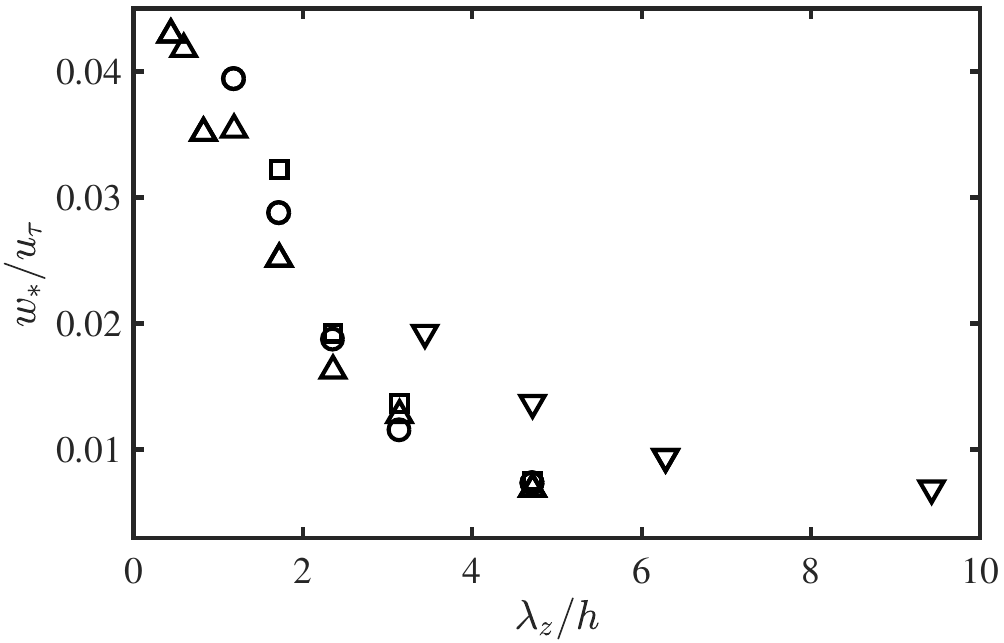}}
\mylab{-.77\columnwidth}{.35\columnwidth}{(d)}%
}%
\caption{%
(a) Premultiplied spectrum of the spanwise velocity versus $\lambda_z$. The
contours are $[0.2,\, 0.8]$ times the overall maximum of all the $\Phi_{ww}^+$.
Dashed and chaindotted lines are the location of the inner and outer maxima of $\Phi_{ww}$ at each
wavelength. Numerical channels \cite{lozano14,lee:moser:15}, with the Reynolds number increasing
from blue to red: $Re_\tau=550,\,1000,\,2000,\, 5200$.
The dashed magenta line is $\lambda_z^+=5y^+$.
(b) Maximum  of $\Phi_{ww}$ versus wavelength, with the same lines as in (a).
(c) Spectral density of the spanwise velocity for various wavelengths from $\lambda_z^+=2\times
10^3$ to $\lambda_z=10h$, from blue to red, integrated in $\lambda_x$ over the diagonal band in Fig
\ref{fig:ww15}, defined as $\lambda_x \in [2,\, 20]\lambda_z$. The symbols are the perturbation
equilibrium profile in Eq. \r{eq:allperturw}, computed from the mean velocity profile in \cite{lee:moser:15}.
(d) Friction velocity perturbation intensities obtained from (c). 
(\dotted and \opentriangledown) Channel \cite{lozano14}, $\retau=550$; the rest are channels
\cite{lee:moser:15}: (\chndot\ and \opensqr) $\retau=1000$; (\dashed\ and \opencirc) $\retau=2000$;
(\solid\ and \opentriangle) $\retau=5200$.
}
\label{fig:spec1dw}
\end{figure}

The perturbation expansion is also slightly different from the streamwise component. The first term
in the expression for $G$ in Eq. \r{eq:allpertur} comes from the incremental boundary layer created
by the perturbation of $\utau$, which in this case has to be oriented spanwise. The second term is
the deformation of the existing boundary layer by the change in length scale due to the new $\utau$,
and is missing from $\delta W$, for which no preexisting spanwise flow exists. The fluctuation
equation becomes,
\beq
(\delta W)'= (\delta w_\tau)' F(y^+)=w_* F(y^+),
\la{eq:allperturw}
\eeq
where $F(y)$ is, as before, the mean velocity profile of a regular channel, and the symbols
$w_\tau$ and $w_*$ have been introduced to represent the friction velocity of the spanwise
perturbation flow.

Figure \ref{fig:spec1dw} summarizes, for the spanwise velocity, the same information as Figs.
\ref{fig:spec1d} and \ref{fig:specden} do for $u$. As in the previous case, the relevant result is
that the transition from the inner to the outer peak scales in wall units, at approximately the same
wavelength as for the streamwise velocity, $\lambda_z^+\approx 1000$ (Fig. \ref{fig:spec1dw}a,b).
Figure \ref{fig:spec1dw}(d) shows that the perturbation $w_*$ required to fit the profiles in Fig.
\ref{fig:spec1dw}(c) to Eq. \r{eq:allperturw} scales well in outer units, so that the total energy
also diverges logarithmically. Notice that the fit of Eq. \r{eq:allperturw} to the data in Fig.
\ref{fig:spec1dw}(c) is as good as that in Fig. \ref{fig:specden}, even if the two predicted
profiles are fairly different.

There are some differences between the two velocity components. The $\Phi_{ww}$ spectra in Fig.
\ref{fig:spec1dw}(a) are consistently wider and taller than $\Phi_{uu}$ in Fig. \ref{fig:spec1d}(c).
The outer peak of $\Phi_{ww}$ follows $\lambda_z\approx 10 y$ instead of $\lambda_z\approx 5y$, and
the distinction between inner and outer intensities in Fig. \ref{fig:spec1dw}(b) is much less clear
than in Fig. \ref{fig:spec1d}(d). While the inner and outer peaks of $\Phi_{uu}$ are
often two distinct maxima separated in $y$, those of $\Phi_{ww}$ are a single maximum whose
location depends on $\lambda$. There are essentially no bimodal intensity profiles in $\Phi_{ww}$.

The distribution of perturbation friction velocities is also different in the two cases. Although
both scale in outer units, the distribution in Fig. \ref{fig:specden}(b) is fairly smooth, with what
appears to be a definite limit at $\lambda_x=0$, but Fig. \ref{fig:spec1dw}(d) could have a more
singular limit at that point.
  
\section{Discussion and conclusions}\la{sec:conc}

In summary, we have seen that, as most things in turbulence, the near-wall region is a multi-scale
flow involving wide ranges of length and width. In consequence, the near-wall peak of the stream-
and spanwise velocity fluctuation intensities cannot be modeled as an elemental object. There is a
viscosity-dominated spectral core, centered at $\lambda^+_x\times\lambda^+_z\approx 1000\times 100$,
which embodies the classical turbulence cycle and scales well across Reynolds numbers, and a
large-scale component that behaves very differently. The high-Reynolds-number limit of the energy
depends on those large scales, which we have explored using spectra. It turns out that they behave
near the wall as internal equilibrium boundary layers, \notyet{driven by the outer
large eddies \cite{bradshaw67,jim18}}. Although this means that they are not essentially different from the
mean flow, and share with it the scaled mean profile and physical parameters, they mimic oscillating
Stokes layers when they are expressed as fluctuations in a Fourier expansion, and become part of
what is known at low Reynolds numbers as the near-wall fluctuation peak.

The intensity of these large-scale fluctuations is predicted to increase logarithmically with
$\retau$ when expressed in wall units, at least within the Reynolds number range of available
experiments, but not to remain concentrated near the wall. At extremely large Reynolds numbers, they
should spread to a fraction of the channel thickness, of the order $O(h/5)$. We have shown that
their profile can be extracted at laboratory Reynolds numbers from the vertical structure of the
spectrum at particular wavelengths. When extrapolated to geophysical Reynolds numbers, they result
in a fairly different fluctuation profile, interestingly including a second maximum away from the
wall.

In fact, the natural consequence of a model in which most of the near-wall kinetic energy at high
Reynolds number is due to interactions with the outer flow is that even this near-wall region, and
any part the boundary layer whose distance from the wall is fixed in wall units, should be
considered as directly driven by the outer flow. Under those circumstances, and even if the velocity
scale for the active turbulence motions responsible for the Reynolds stresses at a given distance
from the wall continues to be the friction velocity, a more natural unit for the integrated inactive motions
may be some measure of the driving velocity ($U_\infty$, $U_{bulk}$ or some velocity combination
such as the mixed scaling in Ref. \onlinecite{deGraaf00}). It is interesting to note that, if we
recall from Eq. \r{eq:loglaw} that $U_\infty\sim \log \retau$, even a logarithmic growth of
$u'^{2+}_p$ implies that $u'_p/U_\infty \sim \log^{-1/2} \retau \to 0$ as $\retau\to \infty$. The
nature of the singularity is not that $u'_p$ tends to infinity, but that $\utau$ tends to zero faster than
$u'_p$.

\acknowledgments{%
This work was supported by the European Research Council under the Caust grant ERC-AdG-101018287. I
am grateful to R. Deshpande, G. Kunkel, M.K. Lee and I. Marusic for the use of their original data,
and to S. Pirozzoli for his thoughtful critique of an early version of this manuscript.
}
 

%
\end{document}